\documentclass[dvips]{article}

\usepackage{icrc2011}

\title{VHE Observations of the Binary Candidate HESS J0632+057 with H.E.S.S. and VERITAS}

\newcommand{\etal}{\MakeLowercase{\textit{et al. }}} 
\shorttitle{G.Maier and J.Skilton \etal VHE Observations of HESS J0632+057}

\authors{Gernot Maier$^{1}$ for the VERITAS Collaboration$^2$ \\ 
Joanna Skilton$^{3}$ for the H.E.S.S. Collaboration}
\afiliations{$^1$DESY, Platanenallee 6, 15738 Zeuthen, Germany \\
$^2$ see Holder et al (these proceedings) for a complete list of authors \\
$^3$  Max-Planck-Institut f\"ur Kernphysik, PO Box 103980, 69029 Heidelberg, Germany}
\email{gernot.maier@desy.de \ \ \ joanna.skilton@mpi-hd.mpg.de}

\abstract{
HESS J0632+057 is an unidentified gamma-ray source located in the Monoceros region, probably  
 associated with the massive Be star MWC 148.
H.E.S.S. and VERITAS observations in the very high energy (VHE) range combined with Swift X-ray data indicate 
that this object is a new member of the elusive gamma-ray binary class. 
We present here results of VHE gamma-ray observations from VERITAS and HESS at energies above 100 GeV taken over more than six years.
The observations confirm HESS J0632+057 as a  point-like VHE source with a significance of more than 12 standard deviations.
The VHE gamma-ray results  are discussed in the context of contemporaneous X-ray observations with Swift XRT.
}

\begin{document}
\maketitle

\section{Introduction}

HESS J0632+057 is a point-like VHE gamma-ray source discovered by the High Energy Stereoscopic System (H.E.S.S.) during observations of the Monoceros Loop supernova remnant in 2004 and 2005 \cite{Aharonian-2007}.
The reported flux of gamma rays with energies above 1 TeV corresponds to about 3\% of the flux of the Crab Nebula,
 with a differential photon spectrum consistent with a power- law function with index of $2.53\pm0.26_{stat}\pm0.20_{sys}$.
The non-detection of HESS J0632+057 by VERITAS during observations in 2006, 2008 and 2009 revealed VHE gamma-ray variability by recording flux upper limits significantly below the 2004-2005 H.E.S.S. fluxes at energies above 1 TeV.

Follow-up X-ray observations with XMM-Newton \cite{Hinton-2009} and radio observations \cite{Skilton-2009} revealed a variable X-ray (XMMU J063259.3+054801) and radio source
with a position compatible with HESS J0632+057 and the B0pe-star MWC 148 (HD 259440).
Long-term X-ray observations using the Swift X-ray Telescope (XRT)
revealed flux modulations with a  period of $321\pm5$ days (see Figure \ref{fig:lightcurve} top), 
which was reported by  \cite{Bongiorno-2011}  as confirmation that HESS J0632+057 is of a binary nature.

The VHE binary class contains currently three established members only: PSR B1959-63/SS 2883, LS 5039 and LS I +61 303.
All show variable emission of gamma rays, which  for PSR B1959-63 and LS 5039 is clearly connected
to the orbital movement of a compact object around a O- or Be-type star (the situation for LS I +61 303 is much less clear, see \cite{Maier-2011}).
The gamma-ray production processes and the location of emission are not known in any of these binaries.
In a general picture, VHE photons are produced by leptonic and hadronic
interactions of high-energy charged particles accelerated in the
shocks generated by the collision of a pulsar wind with the wind of the massive star or, alternatively, in shocks
associated with an accretion-powered jet.
The orbital movement leads to regular changes of physical parameters, thus leading to variability of emission at the site and variability of absorption of that emission.  
This enables studies of the relevant processes in detail under repeating conditions

A detailed modeling of the VHE emission from HESS J0632+057 is hampered by the unknown orbital 
parameters and nature of the compact object.
More data is clearly needed to understand the physics in this system.
We present in the following results from multi-year observations of HESS J0632+057 with the H.E.S.S. 
and VERITAS experiments.

\section{H.E.S.S. and VERITAS}

The High Energy Stereoscopic System (H.E.S.S.) consists of four 13\,m imaging atmospheric-Cherenkov telescopes (IACTs) positioned in a square of side length 120\,m. The array is located in the Khomas highlands of Namibia, 1800m above sea level. The total field of view of the system is 5$^{\circ}$ (details about the detectors can be found in \cite{Aharonian-2006} and references therein). 

VERITAS is an array of four imaging atmospheric-Cherenkov telescopes (IACT)
located at the Fred Lawrence Whipple Observatory
in southern Arizona.
The field of view of the VERITAS telescopes is 3.5$^{\mathrm{o}}$. 
For more details on the VERITAS
instrument, see e.g.~\cite{Holder-2011}.

Both instruments are very similar in their performance with 
 large effective
areas ( $>10^5$ m$^2$) over a wide energy range (100 GeV
to 30 TeV),  as well as good energy (15-20\%) and angular
($\approx0.1^{\mathrm{o}}$) resolution.
The high sensitivity of H.E.S.S. and VERITAS enable
the detection of sources with a flux of 1\% of the Crab
Nebula in less than 30 hours of observations.

\section{Observations of HESS J0632+057}

\begin{figure*}[th]
  \centering
  \includegraphics[width=0.8\linewidth]{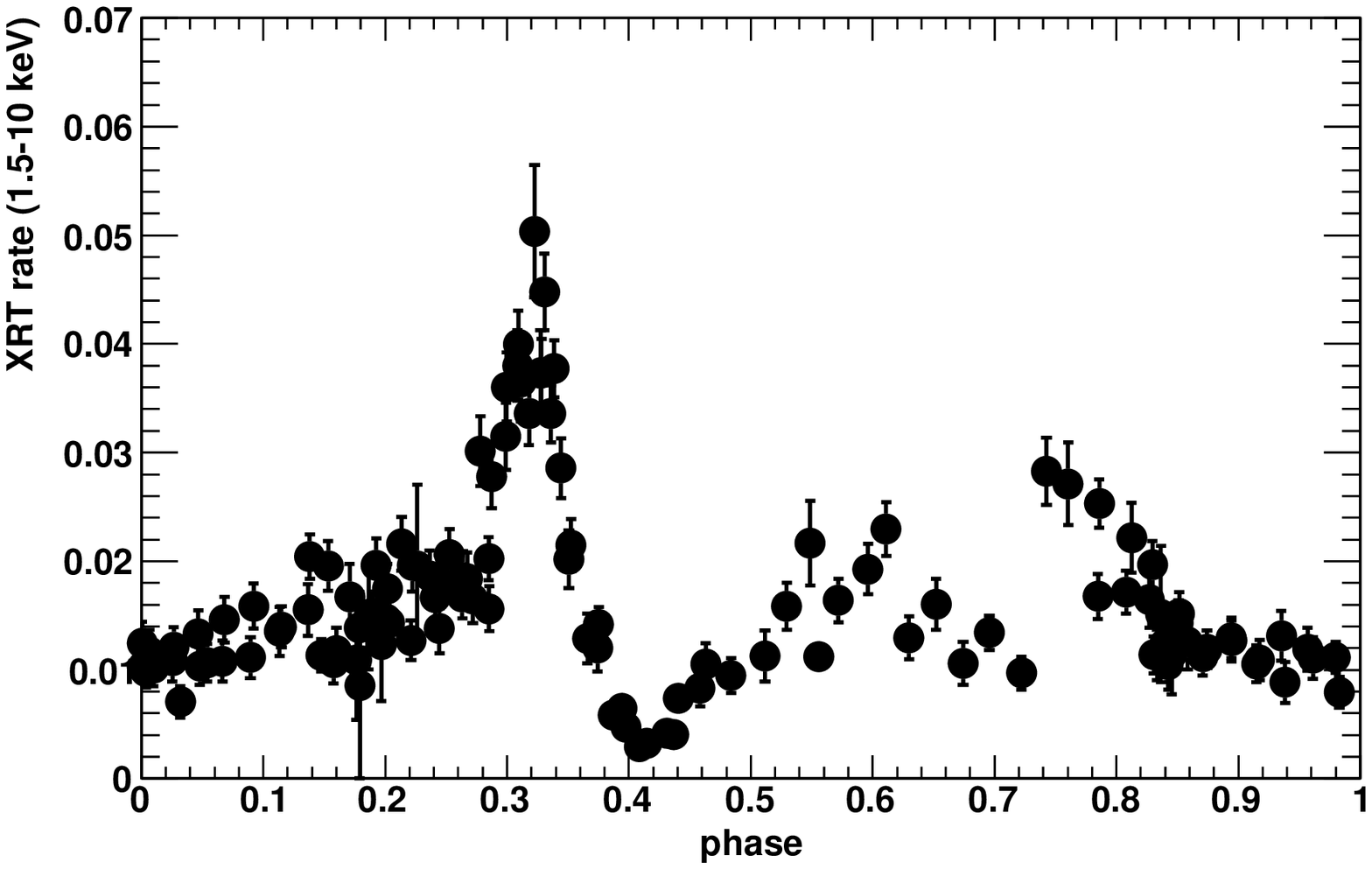}
  \includegraphics[width=0.8\linewidth]{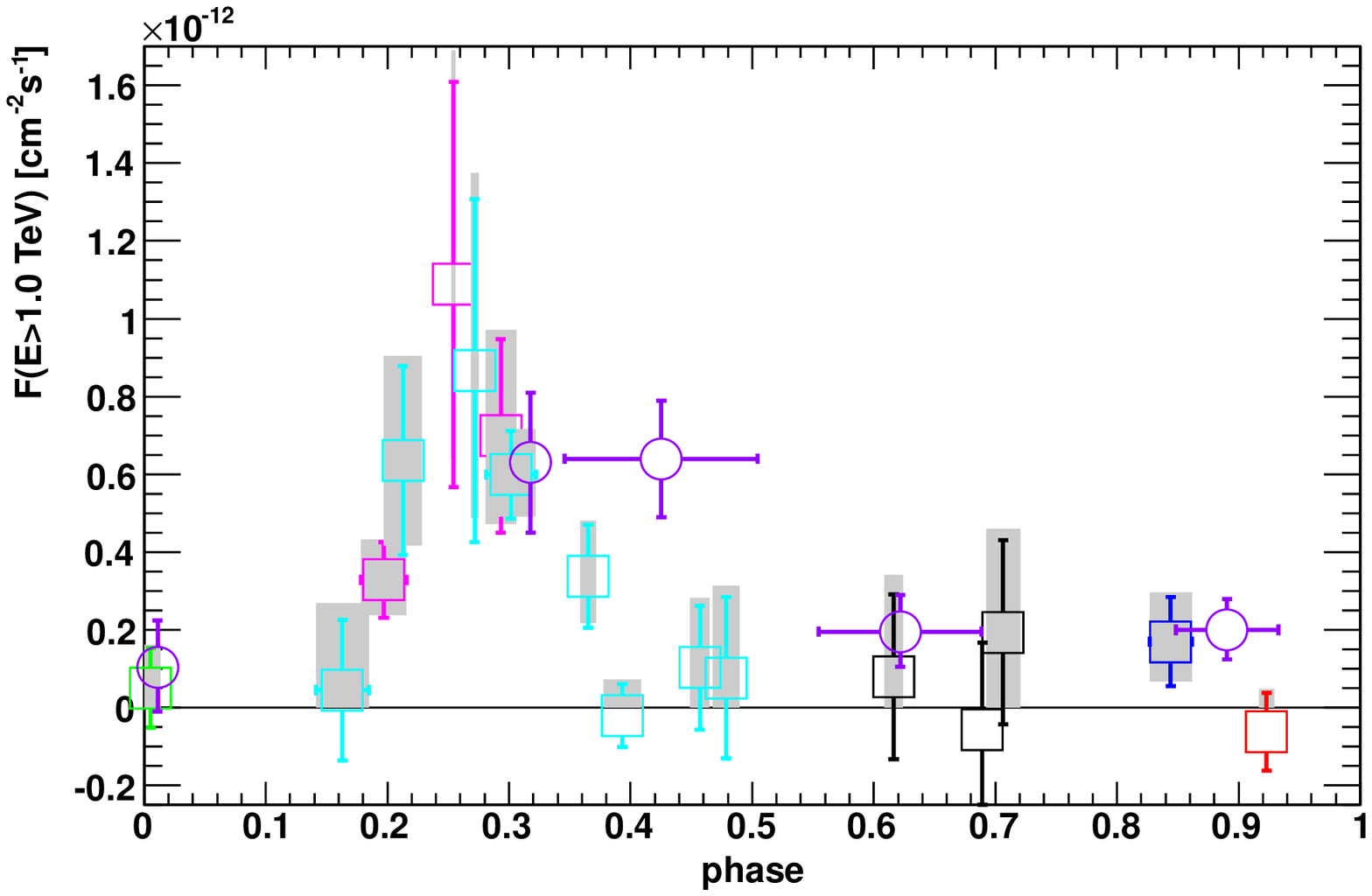}
  \caption{
Top:  Folded X-ray light curve for HESS J0632+057 (XMMU J063259.3+054801) assuming a period of 321 days 
  \cite{Bongiorno-2011} for an energy range 1.5-10 keV. 
  These Swift X-ray observations of the source range from MJD 54857 to 55647, zero phase has been arbitrarily defined
   at MJD 54877, as in \cite{Bongiorno-2011}.
 Bottom:
 VHE observations of HESS J0632+057  folded with the same phase as the top figure.
 Shown are the H.E.S.S. (circular markers) and VERITAS (open squares) measurements as listed in Table \ref{table:observations}.
 The colors of the markers indicate the different  periods of the VERITAS observations 
 (black: Dec 16 2006 - Jan 25 2007, red: Dec 30 2008 - Jan 03 2009, green: Jan 26 2009 - Jan 30 2009,
 purple: Feb 7 2010 - Mar 21 2010, blue:  Oct 18 2010 - Oct 30 2010, turqoise:  Dec 14 2010 - April 5 2011).
  }
  \label{fig:lightcurve}
 \end{figure*}

HESS J0632+057 has been observed by H.E.S.S. and VERITAS in total for more than 150 hours in the past 7 years.
While the timing of the observations in Autumn 2010 and Spring 2011 were motivated by the expected increase in 
X-ray flux as indicated by Swift XRT observations, the dates of all other observations were motivated by scheduling 
and technical constraints only.
Table \ref{table:observations} and Figure \ref{fig:lightcurve} give an overview of the data sets, observing conditions and results.

\begin{table*}[!h]
\centering
 \begin{tabular}{|l|c|c|c|}
 \hline
 Observatory & Date range &  Elevation  & Obs.   \\
            &                        &     range      & time    \\
            &                        &                      & [min] \\
 \hline
 \hline
H.E.S.S.  & Dec 2004 &  31-61$^{\mathrm{o}}$  &280 \\
H.E.S.S.  & Nov 2005 - Dec 2005 & 31-61$^{\mathrm{o}}$ &  370 \\
VERITAS  & Dec 16 2006 - Jan 25 2007  & 55-65$^{\mathrm{o}}$  & 580  \\
H.E.S.S.     & Jan 17 2007 - Mar 20 2007  & 54-62$^{\mathrm{o}}$ & 387   \\ 
H.E.S.S.     & Nov 10 2007 - Jan 10 2008  & 50-62$^{\mathrm{o}}$ & 364   \\ 
VERITAS  & Dec 30 2008 - Jan 03 2009  & 59-65$^{\mathrm{o}}$  & 560  \\
VERITAS  & Jan 26 2009 - Jan 30 2009  & 59-65$^{\mathrm{o}}$  & 720   \\
H.E.S.S.     & Jan 28 2009 - Mar 20 2009  & 40-61$^{\mathrm{o}}$ & 158   \\ 
H.E.S.S.     & Oct 17 2009 - Dec 20 2009  & 40-61$^{\mathrm{o}}$ & 1098    \\ 
VERITAS & Feb 7 2010 - Mar 21 2010 & 53-65$^{\mathrm{o}}$  & 1235  \\
VERITAS & Oct 18 2010 - Oct 30 2010 & 59-65$^{\mathrm{o}}$  & 490  \\
VERITAS & Dec 14 2010 - April 5 2011 & 46-65$^{\mathrm{o}}$  & 2900\\ 
   \hline
  \end{tabular}
  \caption{
   \label{table:observations}
  Details of the H.E.S.S.  and VERITAS observations of \mbox{HESS J0632+057}.
}
  \end{table*}

 \begin{figure*}[th]
  \centering
  \includegraphics[width=0.48\linewidth]{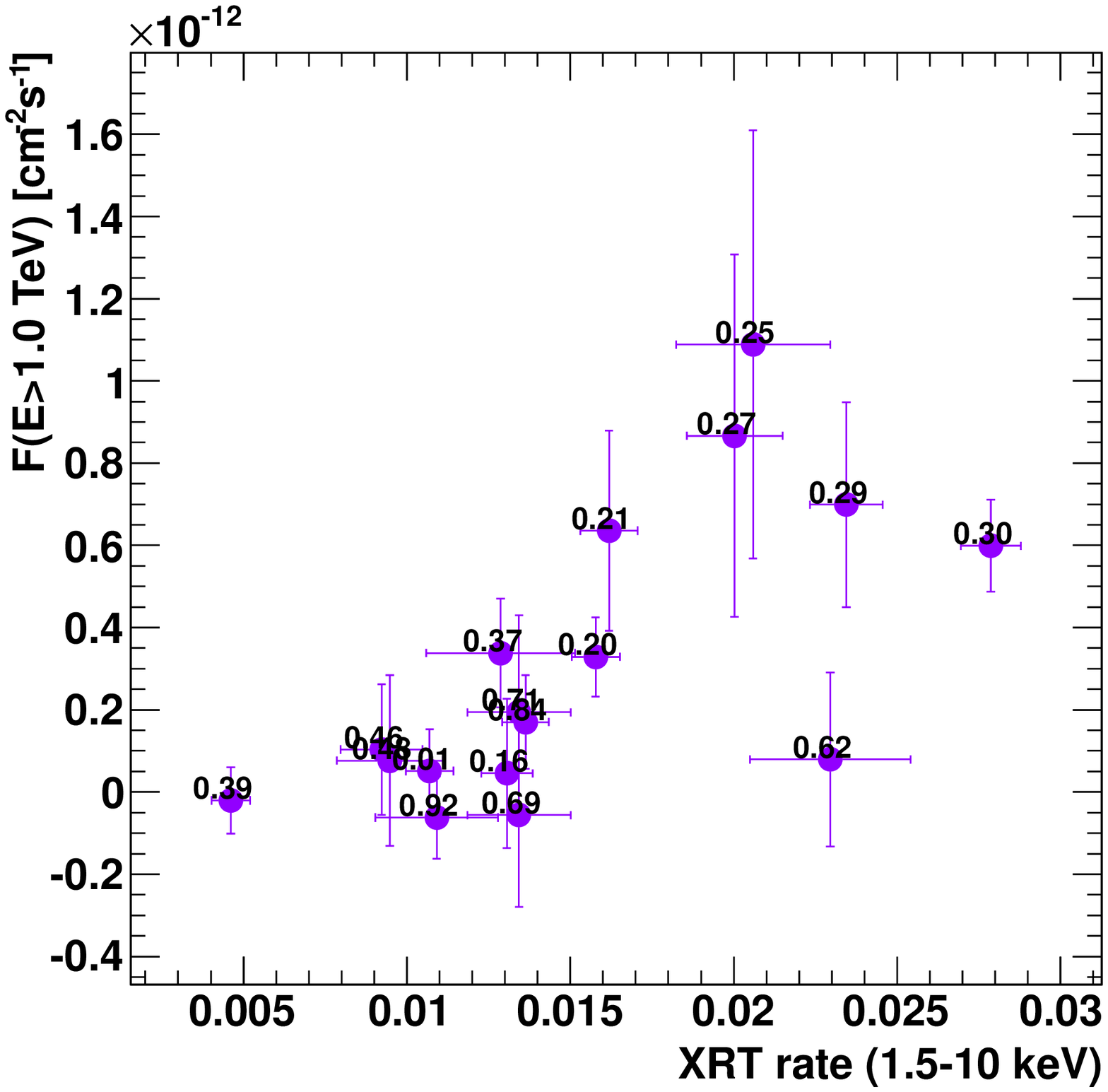}
  \includegraphics[width=0.48\linewidth]{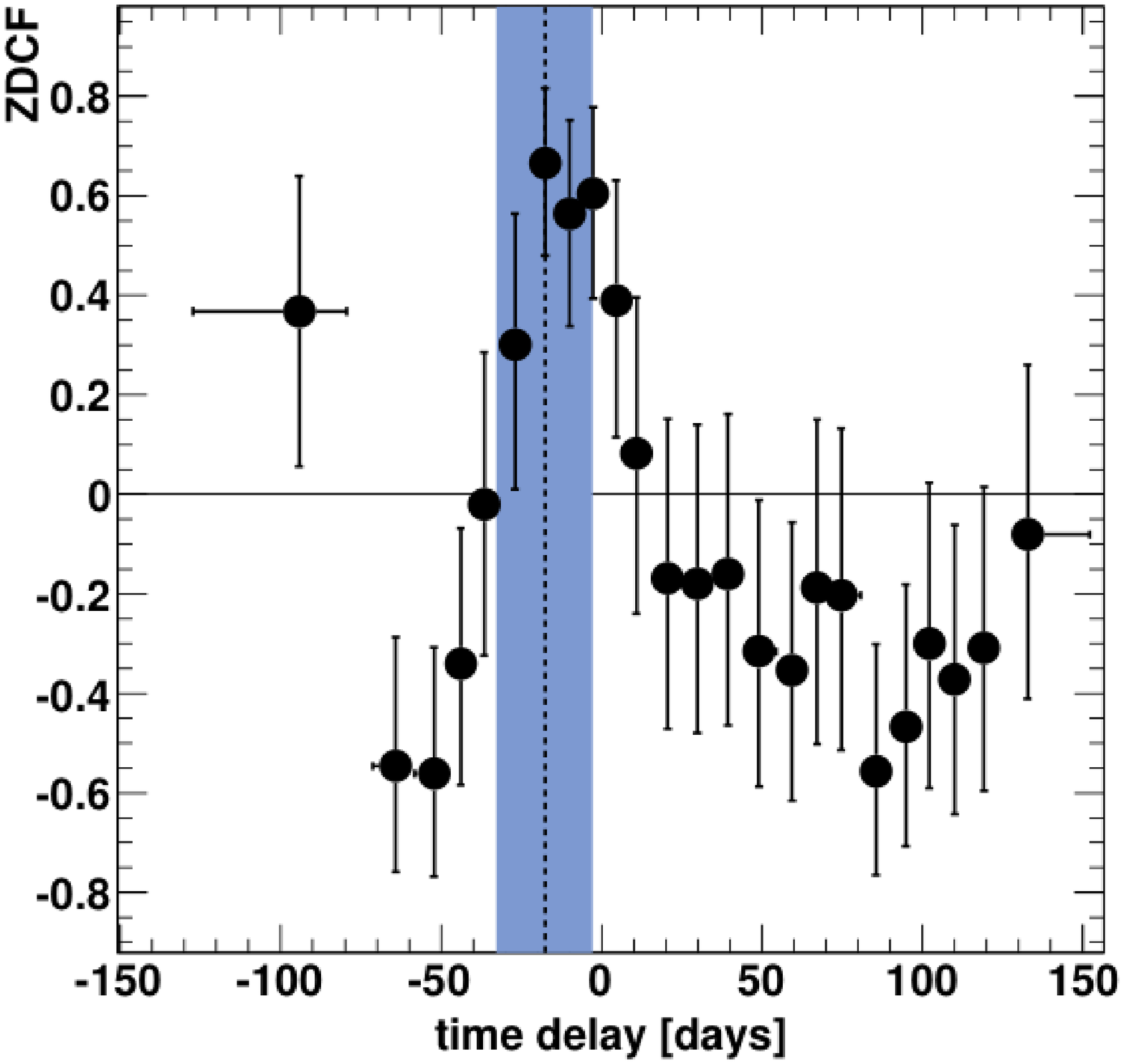}
  \caption{
  Left: VHE gamma-ray vs X-ray fluxes for the same data as presented in Figure \ref{fig:lightcurve}. 
  The X-ray data was averaged over $\pm$1.5 days around the times of the VHE observations. 
  The numbers beside the data points indicate the phases as defined in Figure Figure \ref{fig:lightcurve}. 
  Right: Results from a z-transformed discrete cross-correlation function analysis \cite{Alexander-1997}
  of the X-ray and VHE gamma-ray fluxes. 
  The blue area indicates the 1-sigma range for the time lag between the two regimes
  (-$17.6^{+14.6}_{-15.2}$ days).
 }
   \label{fig:correlations}
 \end{figure*}

VERITAS observed the sky around HESS J0632+057 during five periods between 2006 and 2011.
About 110 hours of observations passed quality selection criteria, which remove data taken during bad weather
or with hardware related problems, see Table \ref{table:observations}.
Data were taken under moonless or moderate moonlight conditions. 
The data sets from prior to 2008 consist of observations during the construction phase of VERITAS with 
three telescopes only.
More than 95\% of data taken from 2008 onwards are taken with the full VERITAS array of four telescopes.
The data analysis steps are described in detail elsewhere (e.g. \cite{Acciari-2009}).

HESS\,J0632+057 was observed yearly with HESS from 2004 until 2009, see Table \ref{table:observations}. The observations were performed over a large range of zenith angles (28$^{\circ}$--58$^{\circ}$) with an average of 34.2$^{\circ}$. The full HESS data set consists of approximately 45 hours of observations (after quality-selection cuts) and has been analysed here using HAP version 11.02.

\section{Results and Discussion}

The long-term observations of HESS\,J0632+057 with H.E.S.S. and VERITAS reveal clearly the variability of this object 
(see Figure \ref{fig:lightcurve} and Table \ref{table:observations}).
H.E.S.S.  discovered it in 2004 and 2005 with a significance of $5.6\sigma$ \cite{Aharonian-2007}, 
it was followed by several years without any reported signal \cite{Acciari-2009}. 
Deep VERITAS observations resulted in the detection of VHE emission in Spring 2009 and Spring 2010, 
the total significance of the VERITAS detection is $>12\sigma$.
A signal from HESS\,J0632+057 in Spring 2011 was also reported  by the MAGIC telescope  \cite{Mariotti-2011}.

Figure \ref{fig:lightcurve} shows the folded X-ray and Gamma-ray light curve assuming a period of 321 days 
(as in \cite{Bongiorno-2011}).
HESS\,J0632+05 has only been detected at phases between approximately 0.2 and 0.5\footnote{
Note that the flux point measured by H.E.S.S. at phase 0.42 is averaged over a time interval of 51 days.},
the position of the maximum in the VHE light curve is at phases $0.27\pm0.5$.

The X-ray and VHE fluxes have been tested for correlations, see Figure \ref{fig:correlations}. 
The  gamma-ray data is sparse, despite the large amount of observations, and does
not allow to measure a significant correlation. 
The light curves suggest that the gamma-ray emission fades away at the onset of the X-ray high state.
The z-transformed discrete cross-correlation function analysis results in a time lag of -$17.6^{+14.6}_{-15.2}$ days,
consistent with a zero time lag.

The astrophysical implications of these observations will be discussed in a forthcoming publication.

\subsubsection*{Acknowledgement}

\begin{tiny}
H.E.S.S.: The support of the Namibian authorities and of the University of Namibia
in facilitating the construction and operation of H.E.S.S. is gratefully
acknowledged, as is the support by the German Ministry for Education and
Research (BMBF), the Max Planck Society, the French Ministry for Research,
the CNRS-IN2P3 and the Astroparticle Interdisciplinary Programme of the
CNRS, the U.K. Science and Technology Facilities Council (STFC),
the IPNP of the Charles University, the Polish Ministry of Science and 
Higher Education, the South African Department of
Science and Technology and National Research Foundation, and by the
University of Namibia. We appreciate the excellent work of the technical
support staff in Berlin, Durham, Hamburg, Heidelberg, Palaiseau, Paris,
Saclay, and in Namibia in the construction and operation of the
equipment.

VERITAS: This research is supported by grants from the US Department of Energy, the US National Science Foundation, and the
Smithsonian Institution, by NSERC in Canada, by Science Foundation Ireland, and by STFC in the UK. We acknowledge the
excellent work of the technical support staff at the FLWO and the collaborating institutions in the construction and
operation of the instrument.
 G.M. acknowledges support through the Young
 Investigators Program of the Helmholtz Association.

This work made use of data supplied by the UK Swift Science Data Centre at the University of Leicester.
\end{tiny}

\clearpage

\end{document}